\documentclass[12pt]{article}
\usepackage{cite,epsfig,amssymb,amsmath,graphicx,color}

\newcommand{\RNum}[1]{\uppercase\expandafter{\romannumeral #1\relax}}

\newcommand{\be}{\begin{equation}}
\newcommand{\ee}{\end{equation}}
\newcommand{\bear}{\begin{eqnarray}}
\newcommand{\eear}{\end{eqnarray}}
\newcommand{\ba}{\begin{array}}
\newcommand{\ea}{\end{array}}
\newcommand{\nn}{\nonumber}

\usepackage[normalem]{ulem}
\usepackage{varioref}
\begin{document}


\vspace{14mm}
\begin{center}
	\Large\bf Hamiltonian Formalism of Topologically Massive Electrodynamics
\end{center}
\hspace{\fill}
\begin{center}
{\large
Taegyu Kim, Seyen Kouwn and Phillial Oh\\[3mm]

{\it Department of Physics and Institute of Basic Science,\\
	Sungkyunkwan University, Suwon 440-746, Korea\\[2mm]

{\tt taegyukim@skku.edu,~seyenkouwn@gmail.com,~ploh@skku.edu}}}

\end{center}

\hspace{\fill}
\vspace{5mm}
\begin{flushleft}
\bf{Abstract}
\end{flushleft}
 
We consider  the   four dimensional topologically massive electrodynamics in which a gauge field is interacting with 2nd rank antisymmetric tensor field through a topological interaction.  The photon becomes massive by eating the 2nd rank tensor field, which is dual to the Higgs mechanism. We explicitly demonstrate 
the nature of the mechanism by performing
a canonical analysis of the theory and discuss various aspects of it.

\vspace{4mm}


\newpage

\section{Introduction}
Massive photon is an old idea \cite{Okun:2006pn} that started with de Broglie who noticed that
photon mass would lead to a faster speed of light with a shorter wavelength. Schr\"odinger point out the exponential cut off of the Earth magnetic field at distances of the order of the massive photon Compton wavelength. The experimental constraints on the photon mass has considerably
increased over the past several decades, putting upper bounds on its mass. So far, the most stringent upper limit is given by $m_\gamma \leq 10^{-27}~{\rm eV}$ \cite{Tu:2005ge}.


On the theoretical side, there are several approaches to endow  photon with mass. The oldest idea is to consider a massive vector field  well-known as Proca theory{}. The gauge invariance can be restored  by introducing a scalar field in the Proca theory \cite{Stueckelberg:1938zz}. Another attempt  is to consider higher derivative Maxwell theory maintaining the gauge invariance \cite{Podolsky:1942zz}.
The Higgs mechanism can be invoked and it is applied to spontaneous breaking of electromagnetic gauge symmetry, yielding a slightly massive photon compatible with experiments \cite{Suzuki:1988bd}. The photon can  also become topologically massive through the Chern-Simons interaction in $2+1$ dimensional electrodynamics \cite{Deser:1981wh}. There are also literatures treating the massive gauge theory in connection with gravity. In Ref. \cite{Padmanabhan:1986yc}, a gauge theory is constructed maintaining the invariance of the action of massless
scalar field under local conformal transformations.  The gauge field  become massive once the scalar is gauge-fixed to a constant value. In Ref. \cite{bogoslovsky}   $R^2$-gravity which has both gauge and conformal invariance is constructed. It is shown that the field equations admit the Proca mass of photon in a suitable gauge.   

In this paper, we consider a topologically massive gauge theory known as $BF$ theory \cite{Cremmer:1973mg}. In this theory, the Maxwell field is interacting with rank two antisymmetric Kalb-Ramond field \cite{Kalb:1974yc} through a topological interaction and it is a gauge theory of massive photon; The gauge field becomes massive by eating the massless Kalb-Ramond field. This alternative to Higgs mechanism has found many applications in diverse  areas of theoretical physics covering from quantum field theory to  superconductivity \cite{Allen:1990gb,Leblanc:1993gx, Deguchi:1999up,Cirio:2013dxa,Choudhury:2015rua,Chatterjee:2016liu,Diamantini:2017fsl}.  Therefore, it is worthwhile to pursue  the  subject from a different perspective. The purpose of this paper is explicit demonstrations of the nature of this mechanism through a canonical analysis  of the topologically massive gauge theory. 

\section{Topologically Massive Photon Model}
We start with the Lagrangian given by \footnote{The conventions for the metric signature and the Levi-Civita
symbol are $(-,+,+,+)$ and $\epsilon_{0123}=1.\,$} 
\bear
{\cal L_{{\rm TMGT}}}&=&{\cal L_{{\rm TM}}} +{\cal L_{{\rm D}}} +{\cal L_{{\rm INT}}},\label{actioon}\\
{\cal L_{{\rm TM}}}&=&-\frac{1}{4}F_{\mu\nu}F^{\mu\nu}
-\frac{1}{12}H_{\mu\nu\rho}H^{\mu\nu\rho}
+\frac{m}{4}\epsilon^{\mu\nu\rho\sigma}B_{\mu\nu}F_{\rho\sigma},\label{tmgt}\\
{\cal L_{{\rm D}}}&=&\bar \psi \gamma^{\mu} i \partial_{\mu}  \psi - m \bar \psi \psi,\nn\\
{\cal L_{{\rm INT}}}&=&A_\mu J^\mu
 +gB_{\mu\nu}\bar J^{\mu\nu}. \label{currentn}
\eear
Here $A_{\mu}$ is the Maxwell field, $B_{\mu\nu}$ is the
antisymmetric Kalb-Ramond tensor field with its field strength defined by
\be
H_{\mu\nu\rho}=\partial_\mu B_{\nu\rho}+\partial_\nu B_{\rho\mu}+\partial_\rho B_{\mu\nu}.
\ee
$m$ is the topological mass of the photon.
$\psi$ is the Dirac fermion and $J^\mu=e\bar \psi \gamma^{\mu} \psi$
is the conserved current. $g$ is a  coupling constant with inverse mass dimension and this model has to be considered as a low-energy effective theory, valid for energy scales below an ultraviolet cutoff $\Lambda$. For the current $\bar J_{\mu\nu}$, several possibilities can be considered \cite{Ogievetsky:1967ij}.
The simplest choice would be $\bar J_{\mu\nu}=0$ \cite{Allen:1990gb}. 
Then, the action (\ref{actioon}) does not contain any dimensionful coupling and it can be  shown that this
theory is a unitary, renormalizable theory of a massive spin-one field with no additional
degrees of freedom. The next possibility is to consider 
\be
\bar J^{\mu\nu}=\frac{g}{2}
\epsilon^{\mu\nu\rho\sigma}\partial_\rho \left(\bar\psi\gamma_\sigma \gamma_5\psi\right),
\ee
in which case, the computation of the lowest-order antisymmetric tensor field contribution
to the muon anomalous magnetic moment and comparison with experiments can be performed \cite{Pilling:2002ij}. As an obvious generalization for
a tensor gauge theory of spin, the standard Dirac tensor
and pseudotensor current densities $\bar J^{\mu\nu}= \bar\psi \sigma^{\mu\nu}\psi$ and $\bar J^{\mu\nu}= \bar\psi i\sigma^{\mu\nu}\gamma^5\psi$ can be considered \cite{Leblanc:1993gx}. Both, however are not suitable for
a gauge theory, since they are not conserved. In this paper, we consider the current given by \cite{Diamantini:2013yka} 
\be
\bar J^{\mu\nu}=-\frac{1}{2}
\epsilon^{\mu\nu\rho\sigma}\partial_\rho J_\sigma, 
\ee
so that ${\cal L_{{\rm INT}}}$ becomes
\be
{\cal L_{{\rm INT}}}=A_\mu J^\mu+\frac{g}{6}\epsilon^{\mu\nu\rho\sigma}H_{\mu\nu\rho}J_\sigma.\label{tint}
\ee
 The action (\ref{actioon})
 is invariant under the gauge transformation
\begin{eqnarray}
\delta A_{\mu}&=&\partial_{\mu}\lambda,~~ \delta\psi= ie\lambda\psi,\nn\\
\delta B_{\mu\nu}&=&\partial_{\mu}\Lambda_{\nu}-\partial_{\nu}\Lambda_{\mu}\label{gt}
\end{eqnarray}
up to a total derivative
term and the gauge invariance is achieved with the current conservation of the Dirac current, $\partial_\mu J^\mu=0$. The interaction preserves $P$ and $T.$ 

The theory described by the Lagrangian ${\cal L_{{\rm TM}}}$ in (\ref{actioon}) could be regarded as either massive gauge theory or
massive anti-symmetric tensor theory. To see that, compute the equations of motions of ${\cal L_{{\rm TM}}}$ which are obtained as
\bear
\partial_\mu F^{\mu\nu} = - \frac{m}{6} \epsilon^{\nu\mu\alpha\beta} H_{\mu\alpha\beta} ,~~ \partial_\mu H^{\mu\alpha\beta}  = -\frac{m}{2} \epsilon^{\alpha\beta\mu\nu} F_{\mu\nu} \ ,
\label{var}
\eear
from which one can derive  \cite{Allen:1990gb}
\begin{equation}
\left[ \Box-m^2 \right] F_{\mu\nu} = 0, ~~~ 
\left[ \Box-m^2 \right] H_{\mu\nu\rho} = 0.
\label{efm}
\end{equation}
The first (second) equation of (\ref{efm}) shows that the fluctuations of the field strength $F_{\mu\nu}$ ($H_{\mu\nu\rho}$) are massive. 
On the other hand, one can eliminate $B_{\mu\nu}$ in the spectrum by considering an auxiliary field formulation.
To show this, let us consider the Lagrangian \cite{Deguchi:1999up}
\begin{eqnarray}
{\cal L}_{\rm A}=-\frac{1}{4}F_{\mu\nu}F^{\mu\nu}
+\frac{m}{4}\epsilon^{\mu\nu\rho\sigma}
B_{\mu\nu} ( F_{\rho\sigma}-2\partial_{\rho}U_{\sigma}) 
-\frac{1}{2}m^{2}U_{\mu}U^{\mu}\label{aux}.
\end{eqnarray}
Here, $U_\mu$ is an auxiliary field which can be eliminated through the equations of motion. We first show that the above action (\ref{aux}) becomes the topologically massive gauge theory ${\cal L_{{\rm TM}}}$  of (\ref{actioon}). The equations of motion for $U_\mu$ give
\begin{eqnarray}
U^{\mu}=-\frac{1}{6m}\epsilon^{\mu\nu\rho\sigma}H_{\nu\rho\sigma}. \label{uhfield}
\end{eqnarray}
%
 Substituting this  into Eq. (\ref{aux}) and
removing a total derivative term, we obtain the Lagrangian ${\cal L_{{\rm TM}}}$.
On the other hand, using the equations of motion for $B_{\mu\nu}$ which read as $\epsilon^{\mu\nu\rho\sigma}\partial_{\rho}(A_{\sigma}-U_{\sigma})=0$,
$U_\mu$ can easily be solved as
\begin{eqnarray}
U_{\mu}=A_{\mu}-\frac{1}{m}\partial_{\mu}\varphi\label{lsol}
\end{eqnarray}
with a scalar field $\varphi$.
Substituting Eq. (\ref{lsol}) into Eq. (\ref{aux}), we obtain the Lagrangian defining the 
St\"{u}ckelberg theory of massive photon:
\begin{equation}
{\cal L}_{\rm S}=-\frac{1}{4}F_{\mu\nu}F^{\mu\nu}
-\frac{1}{2}(mA_{\mu}-\partial_{\mu}\varphi)
(mA^{\mu}-\partial^{\mu}\varphi).\label{st}
\end{equation}
 Thus, using an equation of motion, ${\cal L}_{\rm A}$ reduces to 
${\cal L_{{\rm TM}}}$ or ${\cal L}_{\rm S}$;  The Lagrangians ${\cal L_{{\rm TM}}}$ and ${\cal L}_{\rm S}$ are classically equivalent.  Moreover, their equivalence can also be established at the quantum level by
using the path-integral method \cite{Deguchi:1998xp}. These demonstrate that ${\cal L_{{\rm TMGT}}}$ of (\ref{actioon}) defines
a topologically massive electrodynamics in which a massive photon interacts with Dirac fermions.

It is to be mentioned that the St\"{u}ckelberg field $\varphi$ is a gauge artifact that can be eliminated through gauge fixing. Then, ${\cal L}_{\rm S}$  becomes the Proca theory of massive vector field. The role of 
$\varphi$ becomes apparent when one discuss  unitarity and renormalizability of ${\cal L}_{\rm S}$ at high energy \cite{Aitchison:1980mv}. On the other hand, the mass  of the gauge field $A_\mu$ in ${\cal L}_{\rm S}$ is provided 
by the field strength $H_{\mu\nu\rho}$ through the relation
\be
mA_{\mu}-\partial_{\mu}\varphi=-\frac{1}{6}\epsilon_{\mu\nu\rho\sigma}H^{\nu\rho\sigma},\label{dualrelation}
\ee
coming from (\ref{uhfield}) and (\ref{lsol}), or from  (\ref{var}). 
We can interpret this as follows;  $B_{\mu\nu}$ which has a single degree of freedom due to the gauge invariance (\ref{gt}) is eaten by the gauge field.
This mechanism is similar to the Higgs mechanism, but is fundamentally different in a couple of aspects. First of all, there is no spontaneous symmetry breaking involved. Secondly, it is the antisymmetric tensor field which is  eaten by the gauge field. Since an antisymmetric tensor field is dual to a scalar field classically, this is a dual Higgs mechanism.
From Eq. (\ref{dualrelation}), one can infer that $H_{0ij}$ which is the conjugate momenta of the dual scalar field in the canonical formulation (see below) is the degree of freedom  being incorporated into the longitudinal mode of the massless photon to generate a massive photon. This procedure can be made more transparent and    be demonstrated  succinctly  in the canonical formalism of the dual Higgs mechanism.

In passing, we mention  that ${\cal L_{{\rm TM}}}$ describes also massive Kalb-Ramond theory.
Introducing a new anti-symmetric tensor field via
\be
K_{\mu\nu}=B_{\mu\nu}+\epsilon_{\mu\nu\alpha\beta}\frac{F^{\alpha\beta}}{2m},
\ee
${\cal L_{{\rm TM}}}$ can be expressed as 
\be
{\cal L}_{{\rm MKR}}-\frac{1}{12}H_{\mu\nu\rho}H^{\mu\nu\rho}-\frac{m^2}{4}
B_{\mu\nu}B^{\mu\nu}+\frac{m^2}{4}K_{\mu\nu}K^{\mu\nu}
\ee
The $K$-field can be eliminated via equations of motion and we are left with massive Kalb-Ramond field. 
 
\section{Hamiltonian Analysis}
We  start with a canonical treatment of the  antisymmetric tensor field in order to 
demonstrate the dual Higgs mechanism
in  canonical formalism.
 Let us consider
\bear
{\cal L_{\rm KR}}=
-\frac{1}{12}H_{\mu\nu\rho}H^{\mu\nu\rho}
 \label{kraction}
\eear
Define the conjugate momentum of $B_{ij}\equiv\epsilon_{ijk}Q^k$ and $B_{0i}\equiv C_i$ via
\be
 \pi_{ij}=\frac{\partial{\cal L}_{\rm KR}}{\partial  \dot B_{ij}}=\frac{1}{2}H_{0ij}
\equiv\frac{1}{2}\epsilon_{ijk}P^k, ~~\pi_i=\frac{\partial{\cal L_{\rm KR}}}{\partial  \dot B_{0i}}=0.
 \label{conj}\ee
$C_i $ is a Lagrange multiplier which produce  constraint equations as can be seen in the  canonical form of the Lagrangian (\ref{kraction})
  \be
{\cal L}_{\rm KR}=\vec P\cdot \dot{\vec Q}-\frac{1}{2}
\left\{\vec P^2+(\nabla\cdot \vec Q)^2\right\}-\vec C\cdot (\nabla\times \vec P).\label{KR}
\ee
To perform the constraint analysis, we adopt the method advertised by Faddeev and Jackiw \cite{Faddeev:1988qp} which greatly simplifies the conventional procedure of Dirac's constraint method \cite{Lahiri:1993xv}. Decomposing $\vec P$ and $\vec Q$ into longitudinal($\parallel$) and transverse($\bot$) parts    
\be
 \vec P=\vec P_\bot+\vec P_\parallel, ~~\vec Q=\vec Q_\bot+\vec Q_\parallel,
\label{decom}\ee
which satisfy
\be
 \nabla\cdot\vec P_\bot=0, ~~\nabla\times\vec P_\parallel=0, ~~\nabla\cdot\vec Q_\bot=0, ~~\nabla\times\vec Q_\parallel=0,\label{ms}
\ee
we find the constraint in  (\ref{KR}) can be solved explicitly and eliminates $P_\bot$. Introducing 
$\Vec P=\nabla\theta$ and $\nabla\cdot\vec Q\equiv \eta$, we have 
\be
{\cal L}_{\rm KB}=\eta\dot\theta-\frac{1}{2}
\left\{(\nabla\theta)^2+\eta^2\right\}=\frac{1}{2}\Big\{
\dot\theta^2-(\nabla\theta)^2\Big\},\label{SKRR}
\ee
where the equation of motion of $\eta$ was used in the second equality.
Note that the Lagrangian (\ref{SKRR}) is precisely that of massless scalar field
and we have shown the equivalence of antisymmetric tensor field and 
massless scalar field in the canonical formalism \cite{Freedman:1980us}. $\theta$ is the longitudinal part of the conjugate momentum of the original
antisymmetric tensor field, and in the dual Higgs mechanism, 
it is eaten  by the massless photon to produce massive photon through the topologically interacting $B\tilde F$ term in the action (\ref{actioon}), as we now demonstrate.

Let us consider the action ${\cal L}={\cal L_{{\rm TM}}}  +{\cal L_{{\rm INT}}}$ in (\ref{actioon}) and 
along with (\ref{conj}),
we introduce
\be
\Pi_i=\frac{\partial{\cal L}_{\rm S}}{\partial  \dot A_i}=
F_{0i}=-\vec E. \label{econj}
\ee
Note that $\vec P$ and $-\vec E$ are different from the canonical conjugate momenta defined by $\Pi=\frac{\partial{\cal L}}{\partial  \dot \Phi}, ~\Phi=(B_{ij}, A_i).$ We work with $\vec P$ and $-\vec E$, because use of these variables along with $\vec A$ and $\vec Q$   exhibits more conveniently how $\vec P$ and $\vec Q$ are eventually eliminated from the spectrum and being incorporated into the longitudinal modes of $\vec E$ and $\vec A$.     
The conjugate momentum of $A_0$ is zero as usual and it is a 
Lagrange multipliers.
Then, after a straightforward computations, we find (\ref{actioon}) can be put into the following form:
\bear
{\cal L}&=&{\cal L}_1+{\cal L}_2+{\cal L}_3,\label{adecom}\\
{\cal L}_1&=&-\vec E \cdot \dot{ \vec A}-\frac{1}{2}
\left\{\vec E^2+(\nabla\times \vec A)^2\right\}-
\ A_0\left(\nabla\cdot \vec E-J^0\right)+\vec J\cdot \vec A\label{maxwell}\\
{\cal L}_2&=&\vec P\cdot \dot{\vec Q}-\frac{1}{2}
\left\{\vec P^2+(\nabla\cdot \vec Q)^2\right\}-\vec C\cdot (\nabla\times \vec P)+g(\nabla\cdot\vec Q)J_0-g\vec J\cdot\vec P\label{topmass}\\
{\cal L}_3&=&m\left\{\vec Q\cdot\vec E-
\vec C\cdot (\nabla\times \vec A)\right\},\label{topint}
\eear
The variables can be separated into transverse and longitudinal modes\be
\vec A=\vec A_\bot+\vec A_\parallel,~~ \vec E=\vec E_\bot+\vec E_\parallel,\label{tl}
\ee 
with each mode satisfying the same properties given in (\ref{ms}).
${\cal L}_1$ is the Maxwell theory interacting with external source and let us first check that in the absence of ${\cal L}_2$ and ${\cal L}_3,$ the longitudinal modes can be completely eliminated from ${\cal L}_1.$ Using (\ref{tl}) and writing ${\cal L}_1={\cal L}_{1\bot}+{\cal L}_{1\parallel}$, we find  
\be
{\cal L}_{1\parallel}=-\vec E_\parallel \cdot \dot{ \vec A}_\parallel-\frac{1}{2}
\vec E_\parallel\cdot\vec E_\parallel
+\vec J_\parallel\cdot \vec A_\parallel,\label{llag}
\ee
with the constraint 
\be
\nabla\cdot \vec E=J^0.
\ee
The constraint can be solved explicitly as 
$\vec E_\parallel=\frac{\nabla}{\nabla^2}J^0$. Putting this back into (\ref{llag}) and using the current conservation, we find all that is left is the instantaneous Coulomb interaction
\be
{\cal L}_{1\parallel}=-\frac{1}{2}J^0\left(\frac{1}{-\nabla^2}\right)J^0,\label{coumb}
\ee
up to a total derivative term. 
Note that the elimination of the longitudinal mode  is achieved without appealing to the gauge fixing condition. 

Thus, we find that if ${\cal L}$ of (\ref{adecom}) describes massive photon theory, ${\cal L}_2$ and ${\cal L}_3$ are responsible for the longitudinal modes of the massive photon. In order to proceed, 
 we first rewrite 
Eq. (\ref{topint}) by using (\ref{tl})  together with  (\ref{decom}) as 
 \bear
{\cal L}_3=m\left\{\vec A_\bot \cdot \dot{\vec Q}_\bot+\vec A_\parallel\cdot\vec P_\parallel-
\vec C\cdot (\nabla\times \vec A)-A_0(\nabla\cdot\vec Q)\right\},
\eear
up to  total time derivative terms.
Here, Eq. (\ref{econj}) and 
\be
~\vec A_\parallel \cdot \dot{\vec Q}_\parallel
=\vec A_\parallel\cdot\vec P_\parallel 
\ee
has been used. 

$A_0$ and $\vec C$ are non-dynamical and produce the constraints  given by
\be
\nabla\cdot(\vec E+m\vec Q)=J^0, ~~\nabla\times(\vec P+ m\vec A)=0.
\label{const}
\ee
We use these constraints to eliminate $\vec P_\bot$ and $\vec Q_\parallel$.
From the first equation of (\ref{const}), we have 
\be
\vec Q_\parallel=\frac{1}{m}\Big(\frac{\nabla}{\nabla^2}J^0-\vec E_\parallel\Big),\label{ccons}
\ee
and the second equation of (\ref{const}) yields
\be
\vec P_\bot+m\vec A_\bot=0.\label{tran}
\ee  
The above two relations are prelude to incorporation of $\vec P$ and $\vec Q$ modes into  $\vec A$ and $\vec E$ degrees of freedom.
Substitution of these relations into  (\ref{adecom}) yields
\bear
{\cal L}&=&{\cal L}_\bot+{\cal L}_\parallel,\label{lde}\\ 
{\cal L}_\bot&=&-\vec E_\bot \cdot \dot {\vec A}_\bot-\frac{1}{2}
\left\{\vec E_\bot^2+m^2\vec A_\bot\cdot\vec A_\bot+
(\nabla\times \vec A)^2\right\}+\vec J_\bot\cdot \vec A_\bot
+gm\vec J_\bot\cdot\vec A_\bot,\label{lbot}\\ 
{\cal L}_\parallel&=&-\vec E_\parallel \cdot (\dot {\vec A}_\parallel
-\frac{1}{m} \dot {\vec P}_\parallel)-\frac{1}{2}
\left\{\vec E_\parallel^2+m^2(\vec A_\parallel-\frac{1}{m}\vec P_\parallel)^2
-m^2\vec A_\parallel^2 +\frac{1}{m^2}(\nabla\cdot \vec E-J^0)^2
\right\}\nn\\
&&~~~~~+\vec J_\parallel\cdot (\vec A_\parallel-\frac{1}{m}\vec P_\parallel)-
\frac{g}{m}J^0(\nabla\cdot\vec E-J^0)
-g\vec J_\parallel\cdot\vec P_\parallel,\label{ll}
\eear
where the current conservation $\dot  J^0+\nabla\cdot \vec J=0$ has been used.
Defining  
\be
\vec a_\parallel=\vec A_\parallel-\frac{1}{m}\vec P_\parallel,~~ ~
 \vec b_\parallel=\vec A_\parallel+\frac{1}{m}\vec P_\parallel,\label{eaten}
\ee
$\vec b_\parallel$ becomes an auxiliary field and can be eliminated through the equation of motion given by
\be
\vec b_\parallel=-\vec a_\parallel +2\frac{g}{m}\vec J_\parallel.
\ee
Putting this back into the action (\ref{ll}), we obtain 

\bear
{\cal L}_\parallel&=&-\vec E_\parallel \cdot \dot{ \vec a}_\parallel-\frac{1}{2}
\left\{\vec E_\parallel^2+m^2\vec a_\parallel^2\right\}
+\kappa_+\vec J_\parallel\cdot \vec a_\parallel
+\frac{\kappa_-}{m^2}J^0(\nabla\cdot\vec E)\nonumber\\
&&~~~~~~~~~~~-\frac{1}{2m^2}(\nabla\cdot \vec E)^2
-\frac{2\kappa_{-}-1}{2m^2}(J^0)^2-\frac{g^2}{2}{\vec J_\parallel^{2}},\label{lpar}
\eear
with
\be
\kappa_\pm=1\pm gm.
\ee

${\cal L}_\bot$ of (\ref{lbot}) is already in its canonical form. Without the mass term and the last term, this is the Lagrangian ${\cal L}_1$ of the massless photon with transverse degrees of freedom only. The origin of transverse mass term  $\frac{1}{2}m^2\vec A_\bot\cdot\vec A_\bot$ is  the canonical energy of conjugate momentum of $B_{\mu\nu}$ field, $\frac{1}{2}\vec P^2$ in  (\ref{topmass}) through the constraint (\ref{tran}).
To bring ${\cal L}_\parallel$ of (\ref{lpar}) into the canonical form, we first write
\be
(\vec E_\parallel,~ \vec a_\parallel,~ \vec J_\parallel)=
\nabla ( E_\parallel,~  a_\parallel,~  J_\parallel).
\ee
Then, introducing the canonical variables \cite{Deser:1972wi}
by\bear
\chi&=&-\left( \sqrt{\left(\frac{-\nabla^2}
{m^2}\right)(m^2-\nabla^2)}E_\parallel+
\kappa_- \sqrt{\left(\frac{-\nabla^2}{m^2}\right)
\frac{1}{m^2-\nabla^2}}J^0\right),\nn\\ 
~\phi&=&ma_\parallel \sqrt{\frac{-\nabla^2}{m^2-\nabla^2}},
\label{canvar}
\eear
we have
up to total derivative terms
\bear
{\cal L}_\parallel &=&\chi\dot\phi-\frac{1}{2}\left\{ \chi^2+(\nabla\phi)^2+ m^2\phi^2\right\}+ m \phi \sqrt{\frac{-\nabla^2}{m^2-\nabla^2}}J_\parallel
-g \phi \sqrt{\frac{-\nabla^2}{m^2-\nabla^2}}(2\nabla^2-m^2)J_\parallel
\nn\\
&&-\frac{1}{2}J^0\left
(\frac{(1-gm)^2}{m^2-\nabla^2}\right)J^0+\frac{1}{2}g^2(J^0)^2-\frac{g^2}{2}J_\parallel (-\nabla^2)J_\parallel.\label{longi}
\eear
Note that apart from the terms containing sources, ${\cal L}_\parallel$ is the Lagrangian of a massive scalar field in its canonical form. We have explicitly extracted the longitudinal mode as a massive scalar field. Eq. (\ref{longi})
has well defined limits of both $m$ and $g$ going to zero simultaneously; ${\cal L}_\parallel$ effectively reduces to ${\cal L}_{1\parallel}$ of (\ref{coumb})
with the decoupling of the longitudinal scalar mode. For $g\rightarrow 0$ with a finite $m$, the action reduces to the longitudinal mode of massive Proca field with conserved current \cite{Deser:1972wi}. The interaction between $\phi$ and $J_\parallel$ is suppressed by a factor of $m.$   In the $m\rightarrow 0$ limit with $g$ fixed, the action describes the massless electrodynamics with an additional topological interaction of (\ref{tint}). In this case, the interaction between $\phi$ and $J_\parallel$ is again suppressed by a factor of $g$. For two point charge sources,
$\rho(\vec x, t)=q_1\delta(\vec x-\vec x_1)+q_2\delta(\vec x-\vec x_2),$
the modified Coulomb potential energy  in (\ref{longi})   between the two charges is given by
\be
V=q_1q_2(1-gm)^2\frac{e^{-m\vert\vec x_1-\vec x_2\vert}}{4\pi\vert\vec x_1-\vec x_2\vert }-g^2q_1q_2\delta(\vec x_1-\vec x_2)
\ee
The topological interaction with $g>0$ contributes  to the screening of the Yukawa potential, but $g<0$ contributes oppositely. It also generates an effective topological   point-like interactions between two charges and also longitudinal components of current, $\sim g^{2}\vec J_\parallel\cdot\vec J_\parallel$.    

\section{Conclusion and Discussions}
In summary, we have explicitly demonstrated the nature of dual Higgs mechanism in Hamiltonian analysis. The process works as follows; $\vec P_\bot$ generates  transverse mass term in (\ref{lbot}) via the constraint (\ref{tran}).
$P_\parallel$ is being incorporated into the longitudinal mode of the massless photon in (\ref{eaten}) which can be identified with a massive scalar by non-local field redefinition (\ref{canvar}). It is the the conjugate momenta $\vec P$ of the dual scalar of the antisymmetric tensor  field which is eaten by the gauge field in the dual Higgs mechanism. $\vec Q_\bot$ drops out of the system due to the constraint (\ref{tran}). $Q_\parallel$ was eliminated through the constraint equation (\ref{const}). The advantage of this method is an explicit display of physical degrees of freedom albeit the cost of manifest covariance.
Its canonical form is ready to be quantized with only physical degrees of freedom. It is to be commented that it is obscure to observe how the mechanism is operating  in the conventional Dirac constraint analysis of ${\cal L}_{{\rm TM}}$ of (\ref{tmgt}) \cite{Lahiri:1993xv}.  We also have obtained effective Hamiltonian with various interaction terms which has a well-defined limit of $m$ and $g$ approaching zero. It would be interesting to extend the present analysis to the non-Abelian generalization of two-form field of Ref. \cite{Freedman:1980us}.


The explicit extraction of longitudinal mode of massive photon as a massive scalar field in (\ref{longi}) might have some applications in cosmology. Even though the longitudinal mode decouples from the external sources in the vanishing limit of $m$ and $g$, its gravitational interaction cannot be nullified \cite{Deser:1972wi}. In general, such a massive scalar field with an ultralight mass might have important cosmological implications in relation with dark energy \cite{Frieman:1995pm}; It can act as an effective cosmological constant before relaxing to its minimum value at $\phi=0.$ The merit in the case of massive photon \cite{Kouwn:2015cdw}
is that no extra scalar field is necessary; The longitudinal mode itself is the source of dark energy. This aspect deserves further investigation. 
~\\

\noindent{\bf 
Acknowledgements}\\
\noindent{This work was supported by Basic Science Research Program through the National Research
	Foundation of Korea (NRF) funded by the Ministry of Education (Grant No.
	2015R1D1A1A01056572)(P.O.),
	the National Research Foundation of Korea (NRF) funded by the Ministry of Education (Grant No. NRF-2017R1D1A1B03032970)(S.K.).}




\begin{thebibliography}{99}


\bibitem{Okun:2006pn} 
  L.~B.~Okun,
  Acta Phys.\ Polon.\ B {\bf 37}, 565 (2006)
  [hep-ph/0602036].

\bibitem{Tu:2005ge}
  L.~C.~Tu, J.~Luo and G.~T.~Gillies,
  Rept.\ Prog.\ Phys.\  {\bf 68}, 77 (2005);
  A.~S.~Goldhaber and M.~M.~Nieto,
  Rev.\ Mod.\ Phys.\  {\bf 82}, 939 (2010)
  [arXiv:0809.1003 [hep-ph]].

\bibitem{Stueckelberg:1938zz} 
  E.~C.~G.~Stueckelberg,
  Helv.\ Phys.\ Acta {\bf 11}, 299 (1938).



\bibitem{Podolsky:1942zz} 
  B.~Podolsky,
  Phys.\ Rev.\  {\bf 62}, 68 (1942).
  doi:10.1103/PhysRev.62.68

\bibitem{Suzuki:1988bd}
  M.~Suzuki,
  Phys.\ Rev.\ D {\bf 38}, 1544 (1988).

\bibitem{Deser:1981wh} 
  S.~Deser, R.~Jackiw and S.~Templeton,
  Annals Phys.\  {\bf 140}, 372 (1982)
  [Annals Phys.\  {\bf 281}, 409 (2000)]
  Erratum: [Annals Phys.\  {\bf 185}, 406 (1988)].
  doi:10.1006/aphy.2000.6013, 10.1016/0003-4916(82)90164-6


\bibitem{Padmanabhan:1986yc} 
  T.~Padmanabhan,
  Class.\ Quant.\ Grav.\  {\bf 2}, L105 (1985).
  doi:10.1088/0264-9381/2/5/002

\bibitem{bogoslovsky}
G. Y. ~Bogoslovsky, Class. Quantum Grav. 9, 569 (1992).


\bibitem{Cremmer:1973mg} 
  E.~Cremmer and J.~Scherk,
  Nucl.\ Phys.\ B {\bf 72}, 117 (1974).
  doi:10.1016/0550-3213(74)90224-7


\bibitem{Kalb:1974yc} 
  M.~Kalb and P.~Ramond,
  Phys.\ Rev.\ D {\bf 9}, 2273 (1974).
  doi:10.1103/PhysRevD.9.2273

\bibitem{Allen:1990gb} 
  T.~J.~Allen, M.~J.~Bowick and A.~Lahiri,
  Mod.\ Phys.\ Lett.\ A {\bf 6}, 559 (1991).
  doi:10.1142/S0217732391000580

\bibitem{Leblanc:1993gx} 
  M.~Leblanc, R.~MacKenzie, P.~K.~Panigrahi and R.~Ray,
  Int.\ J.\ Mod.\ Phys.\ A {\bf 9}, 4717 (1994)
  doi:10.1142/S0217751X94001886
  [hep-th/9311076].

  

\bibitem{Deguchi:1999up} 
  S.~Deguchi,
  Phys.\ Lett.\ B {\bf 532}, 329 (2002)
  doi:10.1016/S0370-2693(02)01538-1
  [hep-th/9903135].

\bibitem{Cirio:2013dxa} 
  M.~Cirio, G.~Palumbo and J.~K.~Pachos,
  Phys.\ Rev.\ B {\bf 90}, no. 8, 085114 (2014)
  doi:10.1103/PhysRevB.90.085114
  [arXiv:1309.2380 [cond-mat.str-el]].


\bibitem{Choudhury:2015rua} 
  I.~D.~Choudhury, M.~C.~Diamantini, G.~Guarnaccia, A.~Lahiri and C.~A.~Trugenberger,
  JHEP {\bf 1506}, 081 (2015)
  doi:10.1007/JHEP06(2015)081
  [arXiv:1503.06314 [hep-th]].


\bibitem{Chatterjee:2016liu} 
  C.~Chatterjee, I.~D.~Choudhury and A.~Lahiri,
  Eur.\ Phys.\ J.\ C {\bf 77}, no. 5, 300 (2017)
  doi:10.1140/epjc/s10052-017-4872-z
  [arXiv:1611.06347 [hep-th]].

\bibitem{Diamantini:2017fsl} 
  M.~C.~Diamantini and C.~A.~Trugenberger,
  EPJ Web Conf.\  {\bf 137}, 04002 (2017)
  doi:10.1051/epjconf/201713704002
  [arXiv:1703.08023 [hep-th]].


\bibitem{Ogievetsky:1967ij} 
  V.~I.~Ogievetsky and I.~V.~Polubarinov,
  Sov.\ J.\ Nucl.\ Phys.\  {\bf 4}, 156 (1967)
  [Yad.\ Fiz.\  {\bf 4}, 216 (1966)].

\bibitem{Pilling:2002ij} 
  T.~Pilling, R.~T.~Hammond and P.~F.~Kelly,
  Gen.\ Rel.\ Grav.\  {\bf 36}, 2131 (2004)
  doi:10.1023/B:GERG.0000038627.63711.88
  [hep-th/0205102].

\bibitem{Diamantini:2013yka} 
  M.~C.~Diamantini, G.~Guarnaccia and C.~A.~Trugenberger,
  J.\ Phys.\ A {\bf 47}, 092001 (2014)
  doi:10.1088/1751-8113/47/9/092001
  [arXiv:1310.2103 [hep-th]].

\bibitem{Deguchi:1998xp} 
  S.~Deguchi, T.~Mukai and T.~Nakajima,
  Phys.\ Rev.\ D {\bf 59}, 065003 (1999)
  doi:10.1103/PhysRevD.59.065003
  [hep-th/9804070].

\bibitem{Aitchison:1980mv}
  See I.~J.~R.~Aitchison,
  ``An Informal Introduction to Gauge Field Theories,''
  OXFORD-TP-17-80 for a general introduction.
 

\bibitem{Faddeev:1988qp} 
  L.~D.~Faddeev and R.~Jackiw,
  Phys.\ Rev.\ Lett.\  {\bf 60}, 1692 (1988).
  doi:10.1103/PhysRevLett.60.1692


\bibitem{Lahiri:1993xv} 
  A.~Lahiri,
  Mod.\ Phys.\ Lett.\ A {\bf 8}, 2403 (1993)
  doi:10.1142/S021773239300369X
  [hep-th/9302046].


\bibitem{Freedman:1980us} 
  D.~Z.~Freedman and P.~K.~Townsend,
  Nucl.\ Phys.\ B {\bf 177}, 282 (1981).
  doi:10.1016/0550-3213(81)90392-8
 

\bibitem{Deser:1972wi} 
  S.~Deser,
  Ann.\ Inst.\ H.\ Poincare Phys.\ Theor.\  {\bf 16}, 79 (1972).

\bibitem{Frieman:1995pm} 
  J.~A.~Frieman, C.~T.~Hill, A.~Stebbins and I.~Waga,
  Phys.\ Rev.\ Lett.\  {\bf 75}, 2077 (1995)
  doi:10.1103/PhysRevLett.75.2077
  [astro-ph/9505060].

\bibitem{Kouwn:2015cdw} 
  S.~Kouwn, P.~Oh and C.~G.~Park,
  Phys.\ Rev.\ D {\bf 93}, no. 8, 083012 (2016)
  doi:10.1103/PhysRevD.93.083012
  [arXiv:1512.00541 [astro-ph.CO]].



\end{thebibliography}
\end{document}